\documentclass[10pt,aps,prl,amsmath,amssymb,superscriptaddress,showpacs,floatfix,twocolumn]{revtex4-1}
\pdfoutput=1
\usepackage{bm}
\usepackage{epsfig}
\usepackage[usenames]{color}
\usepackage{graphicx}
\usepackage{hyperref}
\usepackage{esint}
\usepackage{comment}

\usepackage{eurosym}

\newcommand\ii{\mathrm{i}}
\newcommand\ee{\mathrm{e}}
\newcommand\dd{\mathrm{d}}

\newcommand{\beq}{\begin{equation}}
\newcommand{\eeq}{\end{equation}}
\newcommand{\beqa}{\begin{eqnarray}}
\newcommand{\eeqa}{\end{eqnarray}}

\begin{document}

\date{\today}

\title{Resonant tunneling of fluctuation Cooper pairs
}

\author{A.~Galda}
\affiliation{Materials Science Division, Argonne National Laboratory, Argonne, Illinois 60439, USA}
\author{A.\,S.\,Mel'nikov}
\affiliation{Institute for Physics of Microstructures, Russian Academy of Sciences, 603950 Nizhny Novgorod, GSP-105, Russia}
\affiliation{Lobachevsky State University of Nizhny Novgorod, 23 Prospekt Gagarina, 603950, Nizhny Novgorod, Russia}
\author{V.\,M.\,Vinokur}
\affiliation{Materials Science Division, Argonne National Laboratory, Argonne, Illinois 60439, USA}

\begin{abstract}
Superconducting fluctuations have proved to be an irreplaceable source of information about microscopic and macroscopic material parameters  that could be inferred from the experiment~\cite{LV}. According to common wisdom, the effect of thermodynamic fluctuations in the vicinity of the superconducting transition temperature, $T_c$, is to round off all of the sharp corners and discontinuities, which otherwise would have been expected to occur at $T_c$.  Here we report the current spikes due to radiation-induced resonant tunneling of fluctuation Cooper pairs between two superconductors which grow even sharper and more pronounced upon approach to $T_c$.  This striking effect offers an unprecedented tool for direct measurements of fluctuation Cooper pairs' lifetime, which is key to our understanding of the fluctuation regime
. Our finding marks a radical departure from the conventional view of superconducting fluctuations as blurring and rounding phenomenon.
\end{abstract}

\pacs{74.25.N, 74.40.-n, 74.50.+r} 
\maketitle

Does every of the fascinating manifestations of the coherent nature of the Cooper pairs below the superconducting critical temperature have a counterpart in the fluctuation regime above $T_c$? Fluctuation paraconductivity~\cite{Aslamazov, Friedmann}, diamagnetism~\cite{Tinkham, Li10}, Hall and Nernst effects above $T_c$~\cite{Ullah, Serbyn, Xu00, Wang01, Li10}, and many others (see~\cite{LV} and references therein) all come with an affirmative. A fluctuation counterpart of Shapiro steps~\cite{Shapiro}, resonant jumps in dc current in biased Josephson Junctions in response to external ac voltage, foreseen by Kulik~\cite{Kulik}, appears to be of special interest.
We develop a theory of fluctuation Shapiro resonances at temperatures above $T_c$ and find that, contrary to its counterpart below $T_c$, this resonant effect is highly sensitive to the temperature dependence of the lifetime of superconducting fluctuations. We demonstrate that the emergence of a new characteristic time above $T_c$, the lifetime of fluctuation Cooper pairs $\tau_{GL}\sim \hbar/(T-T_c)$, leads to new distinct resonant features in the current-voltage characteristics, which offer a direct access to high-precision measurements of $\tau_{GL}$.

Shapiro resonances that result from the tuning or synchronization between the intrinsic Josephson frequency and that of the external source are one of the profound manifestations of the Josephson physics. The importance of this effect is in that it provides the currently most accurate physical standards for the volt~\cite{Hamilton}.

\begin{figure}[!tbp]
    \includegraphics[width=\columnwidth]{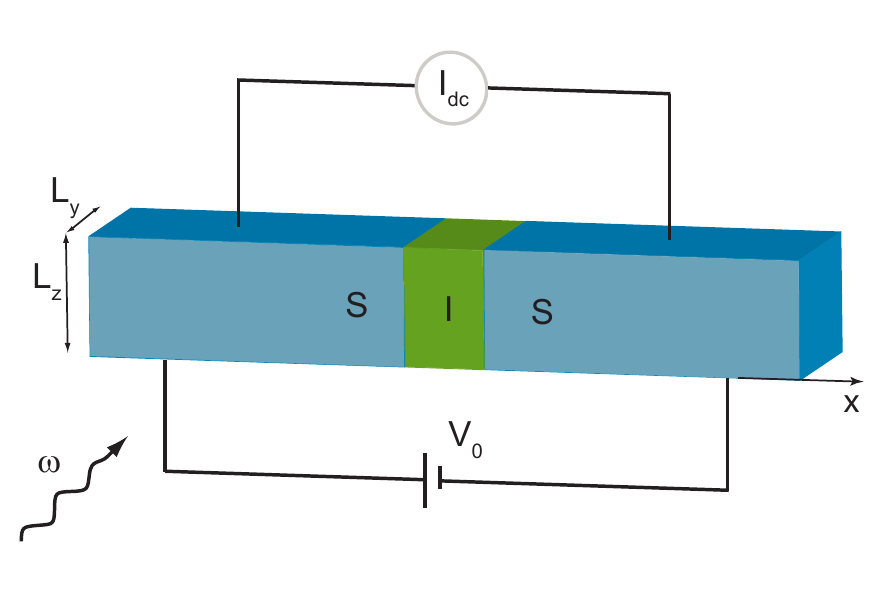}
    \caption{System setup. Proposed 4-point measurement of the IV characteristics of an ac+dc voltage biased tunnel junction. Transition between 1D and 3D geometries can be achieved experimentally by varying the lateral dimensions $L_y$ and $L_z$ of the insulating (I) layer sandwiched between two superconductors (S).}\label{setup}
\end{figure}

Above $T_c$, Shapiro resonances can be most conveniently realized in a biased superconductor-insulator-superconductor (SIS) junction with applied ac voltage induced, e.g., by external microwave radiation of frequency $\omega$, see Fig.~(\ref{setup}). We let the total applied voltage across the junction be $V(t) = V_0 + V_1\cos(\omega t)$.
This produces equally separated jumps in dc current through the junction at the following values of the bias voltage: $V_0^{(n)} = \hbar\omega n/2e$, where $n$ is an integer, i.e. when Cooper pairs with charge $2e$ pass across the junction. To derive the current-voltage (IV) characteristics in Gaussian fluctuation regime we employ the time-dependent Ginzburg--Landau (TDGL) theory linearized in the order parameter~$\psi$:
\begin{align}
    &\left\{-\frac1{4m}\nabla^2 + a + \Gamma^{-1}\left[\partial_t - \ii e V(t)\right]\right\} \psi_R = f(\textbf{r}, t)\,,\label{psi_R}\\
    &\left\{-\frac1{4m}\nabla^2 + a + \Gamma^{-1}\left[\partial_t + \ii e V(t)\right]\right\} \psi_L = f(\textbf{r}, t)\,,\label{psi_L}
\end{align}
where $\psi_L$ and $\psi_R$ are the order parameters in the left ($x<0$) and right ($x>0$) halfspaces,
$a = \alpha (T - T_c) = \alpha T_c\,\epsilon$ and $f(\textbf{r}, t)$ is the Gaussian noise term introduced to satisfy the fluctuation--dissipation theorem, so that
\begin{equation}
    \langle f^*(\textbf{r}, t)f( \textbf{r}^\prime,t^\prime)\rangle = 2T\Gamma^{-1}\delta(\textbf{r} - \textbf{r}^\prime)\delta(t - t^\prime)\,.
\end{equation}
Hereafter we assume $\hbar = k_\text{B} = 1$. Within the standard BCS model the Cooper pair relaxation rate $\Gamma$ is given by the expression: $\Gamma = 8/\pi\alpha$. Linearized TDGL theory describes the system near but certainly outside the range of strong fluctuations defined by the respective Ginzburg criterion~\cite{LV}  $Gi<(T-T_c)/T_c\ll 1$, where $Gi$ is the Ginzburg--Levanyuk number.

The insulating barrier is placed at the plane $x = 0$ and is supposed to be high enough, so that it accommodates most of the voltage drop. We neglect the spatial dependence of the electric scalar potential in the bulk. The width of the insulating layer (the $x$-dimension, see Fig.~(\ref{setup})) is assumed to be small as compared to the Ginzburg--Landau coherence length, in which case the role of the barrier is to impose linear boundary conditions~\cite{deGennes} for the superconducting order parameter in the plane $x = 0$:
\begin{align}
    &\partial_x\psi_R(x=0) - \gamma_1\,\psi_R(x=0) = \beta\,\psi_L(x=0)\,,\label{BC_R}\\
    &\partial_x\psi_L(x=0) + \gamma_2\,\psi_L(x=0) = -\beta\,\psi_R(x=0)\,.\label{BC_L}
\end{align}
These general boundary conditions describe both the suppression of the order parameter at the superconductor-insulator boundary and the Josephson tunneling of Cooper pairs. The constants $\gamma_1$, $\gamma_2$ and $\beta$ depend on the type of material inside the Josephson junction (metal/insulator) and can be determined from a microscopic theory.

The Josephson current through the barrier, averaged over thermal fluctuations, can be written using the above boundary conditions:
\begin{align}\label{current}
    &\langle I(0, t)\rangle = \frac {e}{m}\textit{Im}\left\langle\psi_{L0}^*\partial_x\psi_{L0}\right\rangle
    = -\frac {e\beta}{m}
    \textit{Im}\,\langle\psi_{L0}^*\psi_{R0}\rangle\,,
\end{align}
where $\psi_{L0}$, $\psi_{R0}$ and their derivatives are taken at the barrier plane. The critical current of the junction is related to $\beta$ by the ratio $I_c = e\beta n_s/m$, where $n_s$ is the superfluid density.
Provided the barrier transparency is small, one can evaluate the current within the perturbation theory in dimensionless parameter
$\beta\xi_0$, where $\xi_0 = (4m\alpha T_c)^{-1/2}$ is the Ginzburg--Landau coherence length. Introducing the Fourier transform
$$
\psi_{R,L}(x,\textbf{r}_\parallel,t) =\int\dd\textbf{p}_\parallel\, \psi_{R,L} (x,\textbf{p}_\parallel,t)\ee^{\ii\textbf{p}_\parallel\textbf{r}_\parallel}\,,
$$
one can easily write the following system of coupled integral equations for the wave functions at the barrier plane:
\begin{widetext}
\begin{align}
    \psi_{R0}(\textbf{p}_\parallel) &= \int\limits_{-\infty}^t\!\dd t_1\!\int\limits_0^\infty\!\dd x_1 G_R(0, x_1,\textbf{p}_\parallel; t -
    t_1)f(x_1,\textbf{p}_\parallel, t_1)
     - \frac\beta{4m}\!\int\limits_{-\infty}^t\!\dd t_1\,G_{R}(0, 0,\textbf{p}_\parallel; t - t_1)\psi_{L0}(\textbf{p}_\parallel)\,,\\
    \psi_{L0}(\textbf{p}_\parallel) &= \int\limits_{-\infty}^t\!\dd t_1\!\int\limits_{-\infty}^0\!\dd x_1 G_L(0, x_1,\textbf{p}_\parallel; t -
    t_1)f(x_1,\textbf{p}_\parallel, t_1)
     - \frac\beta{4m}\!\int\limits_{-\infty}^t\!\dd t_1\,G_{L}(0, 0,\textbf{p}_\parallel; t - t_1)\psi_{R0}(\textbf{p}_\parallel)\,,
\end{align}
with the following Green's functions:
\begin{align}
    G_R(x, x',\textbf{p}_\parallel; t - t') &= \sqrt{\frac{m\Gamma}{\pi(t - t')}}\,\ee^{\ii e \int\limits_{t'}^t\!\dd t_1V(t_1) - \tilde a(\textbf{p}_\parallel)\Gamma(t -
    t')}\left[\ee^{-\frac{(x - x')^2}{\frac{\Gamma}m(t - t')}} + \ee^{-\frac{(x + x')^2}{\frac{\Gamma}m(t - t')}} - 2\gamma_1\!\int\limits_0^\infty\!\dd
    y\,\ee^{-\gamma_1 y - \frac{(x + x' + y)^2}{\frac{\Gamma}m(t - t')}}\right]\,,\label{G_R}\\
    G_L(x, x',\textbf{p}_\parallel; t - t') &= \sqrt{\frac{m\Gamma}{\pi(t - t')}}\,\ee^{-\ii e \int\limits_{t'}^t\!\dd t_1V(t_1) - \tilde a(\textbf{p}_\parallel)\Gamma(t -
    t')}\left[\ee^{-\frac{(x - x')^2}{\frac{\Gamma}m(t - t')}} + \ee^{-\frac{(x + x')^2}{\frac{\Gamma}m(t - t')}} -
    2\gamma_2\!\int\limits_{-\infty}^0\!\dd y\,\ee^{\gamma_2 y - \frac{(x + x' + y)^2}{\frac{\Gamma}m(t - t')}}\right]\label{G_L}\,,
\end{align}
where $\tilde a(\textbf{p}_\parallel) = a + p_\parallel^2/4m$. In the limit of weak tunneling current through a symmetric junction, i.e. $\beta\xi_0 \ll 1$ and $\gamma_1 = \gamma_2$,  Eq.~(\ref{current}) yields (up to the second order in $\beta\xi_0$):
\begin{align}\label{J0tau}
    &\langle I(0, \textbf{p}_\parallel; t)\rangle = \frac{\beta^2eT\Gamma^{-1}}{m^2}\textit{Im}\left\{\!\int\limits_{-\infty}^t\!\dd t_1\,G_L^*(0, 0, \textbf{p}_\parallel; t -
    t_1)\!\int\limits_{-\infty}^{t_1}\!\dd t_2\!\int\limits_0^\infty\!\dd x\,G_R^*(0, x, \textbf{p}_\parallel; t_1 - t_2)G_R(0, x, \textbf{p}_\parallel; t - t_2)\right\}\notag\\
    & = \frac{8\beta^2eT\Gamma}{\pi m}\sqrt{\frac{m}{\pi\Gamma}}\!\int\limits_{-\infty}^t\!\frac{\dd t_1}{\sqrt{t - t_1}}\,\textit{Im}\left[\ee^{2\ii e\left\{V_0(t - t_1) + \frac{V_1}\omega\left[\sin(\omega t) - \sin(\omega t_1)\right]\right\}}\right] \left[ 1 - \gamma_1\!\int\limits_{-\infty}^0\!\dd y\, \ee^{\gamma_1 y - \frac{y^2}{\frac{\Gamma}m(t - t_1)}}\right]\notag\\
    &\times \int\limits_{-\infty}^{t_1}\!\dd t_2\,\frac{\ee^{-2\tilde a\Gamma(t - t_2)}}{\sqrt{(t - t_2)(t_1 - t_2)}}\int\limits_0^\infty\!\dd x\left[\ee^{-\frac{x^2}{\frac{\Gamma}m(t_1 - t_2)}} - \gamma_1\!\int\limits_0^\infty\!\dd y\,\ee^{-\gamma_1y - \frac{(x +
    y)^2}{\frac{\Gamma}m(t_1 - t_2)}}\right]\left[\ee^{-\frac{x^2}{\frac{\Gamma}m(t - t_2)}} - \gamma_1\!\int\limits_0^\infty\!\dd y\,\ee^{-\gamma_1y - \frac{(x + y)^2}{\frac{\Gamma}m(t - t_2)}}\right]\,.
\end{align}
It is convenient to use the following identity:
\begin{align}
    &\textit{Im}\left[\ee^{2\ii e\left\{V_0(t - t_1) +
    \frac{V_1}\omega\left[\sin(\omega t) - \sin(\omega t_1)\right]\right\}}\right] = \textit{Im}\left[\ee^{2\ii eV_0(t -
    t_1)}\sum\limits_{k=-\infty}^{\infty}\!J_k\!\left(\frac{2eV_1}\omega\right)\ee^{\ii k\omega t}
    \sum\limits_{k'=-\infty}^{\infty}\!J_{k'}\!\left(\frac{2eV_1}\omega\right)\ee^{-\ii k'\omega t_1}\right]\notag\\
    &= \sum\limits_{k=-\infty}^{\infty}\sum\limits_{k'=-\infty}^{\infty}\!J_k\!\left(\frac{2eV_1}\omega\right)J_{k'}\!\left(\frac{2eV_1}\omega\right)
    \sin\left[2eV_0(t - t_1) + (k - k')\omega t + k'\omega(t - t_1)\right]\,.
\end{align}
The dc component of the current through the junction is found for $k = k'$, while other terms average to zero over one oscillation period $2\pi/\omega$:
\begin{align}
    &\langle I(0, \textbf{p}_\parallel; t)\rangle_\text{dc} = \frac{8\beta^2eT}{\pi}\sqrt{\frac{\Gamma}{\pi m}}\sum\limits_{k=-\infty}^{\infty}\!J_k^2\!\left(\frac{2eV_1}\omega\right)\!\int\limits_{-\infty}^t\!\frac{\dd t_1}{\sqrt{t - t_1}}\,\sin\left[ \left(\frac{2eV_0}\omega + k\right)\left( \frac{\pi\omega}{8T_c}\right) \frac{t - t_1}{\epsilon + \rho}\right]\left[ 1 - \gamma_1\!\int\limits_{-\infty}^0\!\dd y\, \ee^{\gamma_1 y-\frac{y^2}{\frac{\Gamma}m(t - t_1)}}\right]\notag\\
    &\times \int\limits_{-\infty}^{t_1}\!\dd t_2\,\frac{\ee^{-2\tilde a\Gamma(t - t_2)}}{\sqrt{(t - t_2)(t_1 - t_2)}}\int\limits_0^\infty\!\dd x\left[\ee^{-\frac{x^2}{\frac{\Gamma}m(t_1 - t_2)}} - \gamma_1\!\int\limits_0^\infty\!\dd y\,\ee^{-\gamma_1y - \frac{(x +
    y)^2}{\frac{\Gamma}m(t_1 - t_2)}}\right]\left[\ee^{-\frac{x^2}{\frac{\Gamma}m(t - t_2)}} - \gamma_1\!\int\limits_0^\infty\!\dd y\,\ee^{-\gamma_1y - \frac{(x + y)^2}{\frac{\Gamma}m(t - t_2)}}\right]\,.
\end{align}
The main result of this Letter can be captured even in the limit $\gamma_1 = \gamma_2 = 0$, when the above expression takes a simpler form:
\begin{align}\label{IVeq}
    \langle I(0, \textbf{p}_\parallel)\rangle_\text{dc} &= \frac{2\beta^2eT}{\sqrt{\pi}m\tilde a(\textbf{p}_\parallel)} \sum\limits_{k =
    -\infty}^{\infty}\!J_k^2\!\left(\frac{2eV_1}\omega\right)
    \int\limits_0^\infty\!\dd \tau\, \sin\left[\left(\frac{2eV_0}\omega + k\right)\frac{\omega\tau}{\tilde a(\textbf{p}_\parallel)\Gamma}\right]
    \frac{\ee^{-\tau}}{\sqrt{\tau}}\,\rm{erfc}\left(\sqrt\tau\right)\,,
\end{align}
\end{widetext}
where $\rm{erfc}(x)$ is the complementary error function.
In order to calculate the total tunneling current across the junction we must sum the expression~(\ref{IVeq}) over all transverse momenta $\textbf{p}_\parallel$:

\begin{equation}
    \langle I\rangle_\text{dc} = \sum_{\textbf{p}_\parallel} \langle I(0, \textbf{p}_\parallel)\rangle_\text{dc}\,.
\end{equation}

To reveal the dependence of the IV curve on the system dimensionality one can evaluate the current for different
transverse dimensions $L_y\,, L_z $ of the superconducting electrode cross section.
Considering the 3D limit ($L_y\,, L_z \gg \xi_0$) one obtains:
\begin{widetext}
\begin{equation}\label{3D}
    \langle I\rangle^{3D}_\text{dc} = n_f^{3D}I_0 \sum\limits_{k = -\infty}^{\infty}\!J_k^2\!\left( \frac{2eV_1}\omega\right) \int\limits_0^\infty\!\dd \tau
    \frac{\ee^{-\tau}}{\sqrt{\tau}}\,{\rm{erfc}}\left( \sqrt\tau\right) \int\limits_0^1\!\frac{\dd \rho}{\epsilon + \rho}\sin\left[ \left(\frac{2eV_0}\omega + k\right)\left(
    \frac{\pi\omega}{8T_c}\right) \frac{\tau}{\epsilon + \rho}\right]\,,
\end{equation}
\end{widetext}
where $I_0 = 8\beta^2eT\xi_0^2/\sqrt{\pi}$ and $n_f^{3D}~=~L_yL_z/(2\pi\xi_0)^2$ is the effective number of fluctuation modes. The integral over the $\rho$ variable has an upper cut-off of the order unity, since the momenta $p_\parallel$ of the fluctuating wave functions within the Ginzburg--Landau theory should be smaller than the inverse coherence length $\xi_0^{-1}$. By analogy, for a thin film with $L_z \lesssim \xi_0\,, L_y \gg \xi_0$ we find:
\begin{widetext}
\begin{equation}\label{2D}
    \langle I\rangle^{2D}_\text{dc} = n_f^{2D}I_0 \sum\limits_{k = -\infty}^{\infty}\!J_k^2\!\left( \frac{2eV_1}\omega\right) \int\limits_0^\infty\!\dd \tau
    \frac{\ee^{-\tau}}{\sqrt{\tau}}\,{\rm{erfc}}\left( \sqrt\tau\right) \int\limits_0^1\!\frac{\dd \rho}{\epsilon + \rho^2}\sin\left[ \left(\frac{2eV_0}\omega + k\right)\left(
    \frac{\pi\omega}{8T_c}\right) \frac{\tau}{\epsilon + \rho^2}\right]\,,
\end{equation}
\end{widetext}
where $n_f^{2D} = L_y/2\pi\xi_0$.
Finally, the current $\langle I\rangle^{1D}_\text{dc}$ for a thin wire with $L_y, L_z \lesssim \xi_0$ is given by  the  Eq.~(\ref{IVeq}), where one puts $\textbf{p}_\parallel = 0$.

\begin{figure}[!bp]
    \includegraphics[width=\columnwidth]{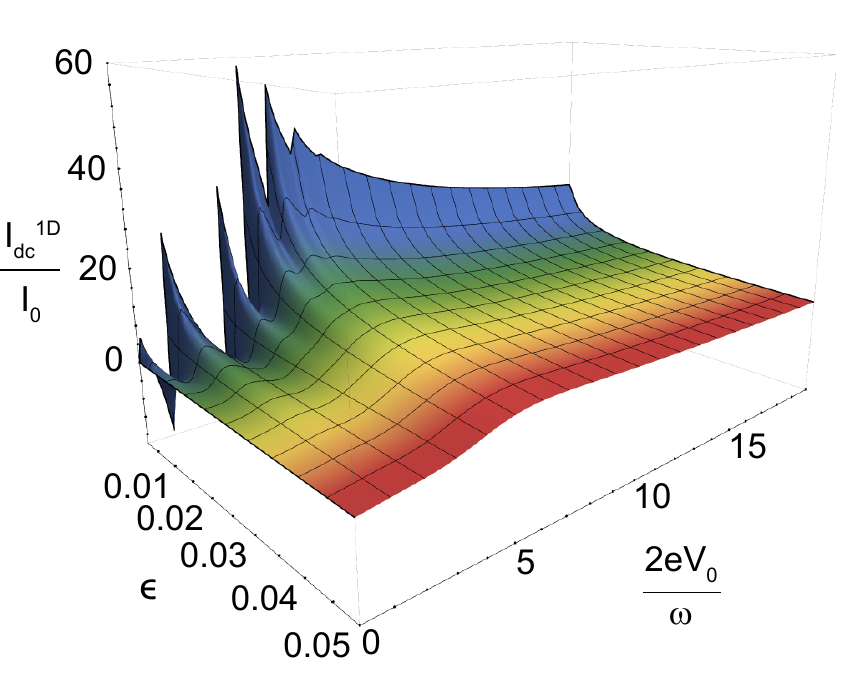}
    \caption{Numerical simulation of IV characteristics as a function of reduced temperature $\epsilon$ and applied dc voltage $V_0$ for a 1D junction at temperatures above $T_c$. Sharp linear spikes in tunneling current at resonant dc voltages $V_0^{(n)}$ are developed upon approaching $T_c$. The corresponding values of $\omega\, \tau_\text{GL}$ are between 2 and 100, which illustrates a transition between the stationary and resonant regimes. Parameters used in the simulation: $2eV_1 = 5\omega$, $\pi\omega/8T_c = 0.1$, which experimentally corresponds to a junction with $T_c = 10$ K under a microwave radiation of frequency $53$ GHz and amplitude $V_1 = 0.33$ mV.}\label{eV0J}
\end{figure}

\begin{figure}[!h]
    \includegraphics[width=\columnwidth]{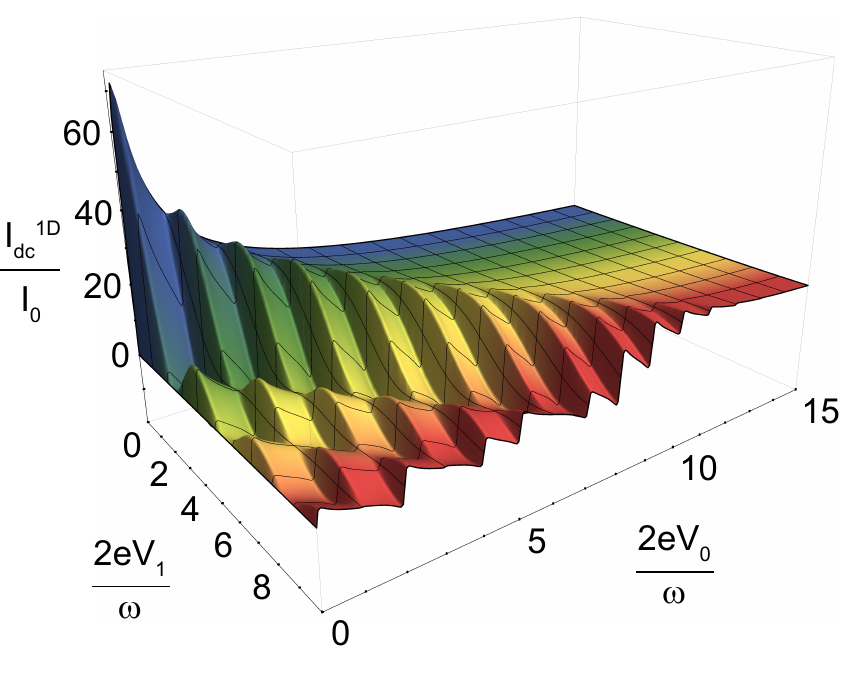}
    \caption{Resonant Shapiro spikes as a function of applied dc and ac voltage. The number of resonances is directly proportional to the amplitude of the applied ac voltage, $V_1$, and is approximately equal to $2eV_1/\omega$. The modulation along the $2eV_1/\omega$ axis is caused by the square of the Bessel function, which explains variable amplitudes of the peaks on IV curves. We took $\epsilon = 0.004$ in this simulation.}\label{V0V1J}
\end{figure}

We first analyze the IV characteristics of a 1D junction. Depending on the parameters of the system, we find that there are three main regimes: 1) $\omega\, \tau_\text{GL} \ll 1$, 2) $\omega\, \tau_\text{GL} \gg 1$ and 3) $\omega\, \tau_\text{GL} \sim 1$, where $\tau_\text{GL} \equiv \pi/8T_c\epsilon$. In the first case, the frequency of the external ac source is so small that temporal variations in applied voltage over the lifetime of Cooper pairs are negligible and the system is in a stationary regime. It follows from our calculations that when a positive constant voltage $V$ is applied to the junction one obtains a smooth IV curve, which starts at the origin, reaches its maximum at $2e V_\text{max} \tau_\text{GL} \simeq 5$ and decays as $\log^{-1}(V)$ at large voltages. Therefore, when averaged over a period $2\pi/\omega$, the peak of the stationary IV characteristic $I(V_0)$ is reached at $V_{0\,\text{max}} \simeq 5/(2e\, \tau_\text{GL})$. The second regime, $\omega\, \tau_\text{GL} \gg 1$, is the primary focus of this Letter. It is characterized by the appearance of equidistant resonant Shapiro spikes on the IV curve at bias voltages $V_0^{(n)} = n \omega/2e$, see Fig.~(\ref{eV0J}).  The number of these resonant features is limited by the value of the ratio $2e V_1/\omega$, with only one peak at the origin when $2e V_1/\omega \ll 1$, see Fig.~(\ref{V0V1J}). The resonances appear on the IV curve as linear spikes with the slope proportional to $\omega\, \tau_\text{GL}/\epsilon$. We note a remarkable universality of the shape of the peaks at fixed $2e V_1/\omega$ and $n$: their height and width are equal to $c_1\tau_\text{GL}$ and $c_2 \tau^{-1}_\text{GL}$, correspondingly, where $c_1$ and $c_2$ are some fixed numeric parameters independent of $\tau_\text{GL}$. This unique property of Shapiro resonances above $T_c$ can serve as a great experimental tool for direct determination of the lifetime of fluctuation Cooper pairs. Finally, when $\omega\, \tau_\text{GL} \!\sim\! 1$ one observes a smooth transition between the stationary and resonant regimes, which occurs when the slope of the IV curves around $V_0^{(n)}$ becomes comparable to the slope of the stationary IV characteristic. We find the same inverse logarithmic decay of the tunneling current at high bias voltages regardless of the value of $\omega\, \tau_\text{GL}$. We also note that external ac voltage can cause Cooper pair fluctuations to tunnel in the \textit{opposite} direction to the applied bias at voltages just below the resonant values $V_0^{(n)}$ for low $n$.

\begin{figure*}[!tbp]
    \includegraphics[width=2\columnwidth]{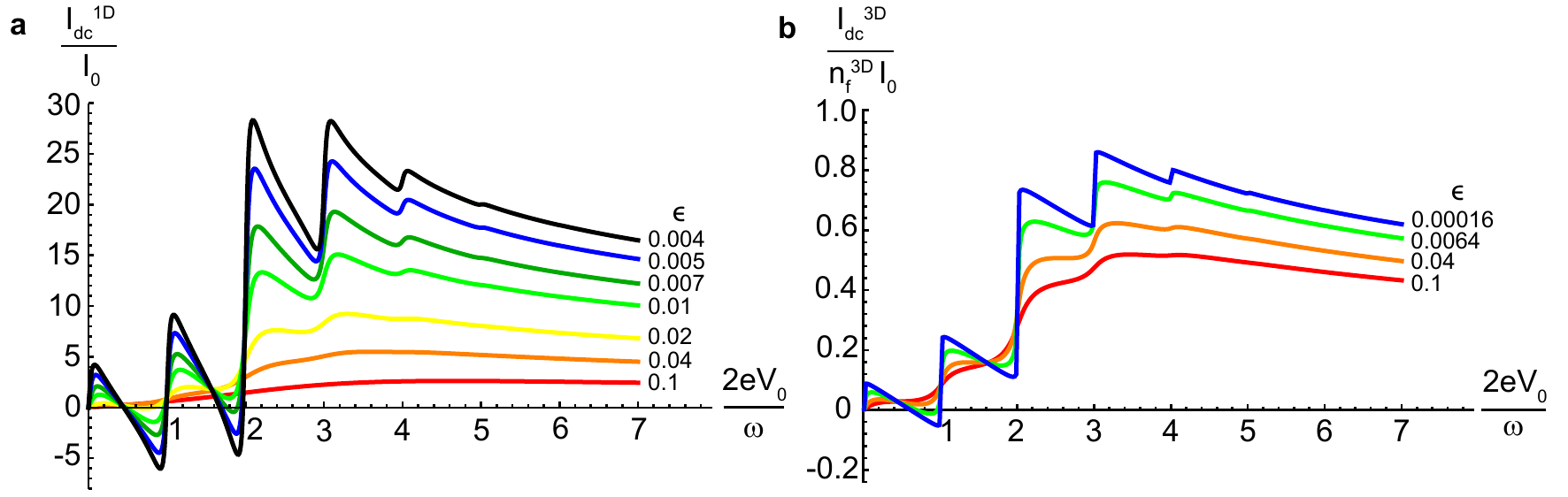}
    \caption{IV characteristics for 1D and 3D SIS junctions in fluctuation regime. \textbf{a}, Tunneling current through a 1D junction normalized by $I_0$ develops Shapiro spikes at resonant values $V_0^{(n)}$ upon approaching $T_c$ due to increasing fluctuation Cooper pair lifetime. At higher dc bias voltages the current decreases monotonically as a function of $V_0$. System parameters used for the calculation: $2eV_1 = 3\omega$, $\pi\omega/8T_c = 0.2$. \textbf{b}, Tunneling current in the 3D model, normalized by $I_0$ and the effective number of fluctuation modes $n_f^{3D}$. For the resonant spikes in the 3D setup to be visible one requires higher ratios of $\omega/T_c$ as compared to 1D system. Parameters used: $2eV_1 = 3\omega$, $\pi\omega/8T_c = 2$, which is equivalent to $T_c = 10$ K, microwave frequency $1$ THz and $V_1 = 6.6$ mV.}\label{J1D_J3D}
\end{figure*}

In Fig.~(\ref{J1D_J3D}) we plot IV characteristics for 1D and 3D setups at different temperatures illustrating, thus, that the effect of resonant fluctuation Cooper pair tunneling should be observable across all effective dimensionalities of the SIS junction. The total current in a 3D case, depicted on Fig.~(\ref{J1D_J3D} b), is normalized by the effective number of fluctuation modes $n_f^{3D}$. As can be seen from the expression for the current in 3D, Eq.~(\ref{3D}), $\omega\,\tau_\text{GL}$ is no longer a single parameter defining the shape of the resonances. Instead, the ratio $\omega/T_c$ determines how well pronounced the spikes around $V_0^{(n)}$ can be upon approaching the critical temperature, i.e. lowering $\epsilon$. In order to clearly observe sharp Shapiro resonances in the Gaussian fluctuation regime for a 3D junction much higher values of $\omega/T_c$ are required as compared to the 1D case, by an order of magnitude or more, assuming the same range of temperatures. This is due to the fact that the slope of Shapiro spikes in 3D is much less sensitive to $\epsilon$ as one approaches $T_c$. In fact, in close vicinity of each resonant value $V_0^{(n)}$ the tunneling current takes the following form:
\begin{equation}
    \langle I \rangle_\text{dc} \propto J_n^2\left( \frac{2eV_1}{\omega}\right)\omega\,\tau_\text{GL}\,\epsilon^{\frac{d - 3}2} \left(V_0 - V_0^{(n)}\right)\,,
\end{equation}
where $J_n$ is the Bessel function of the first kind and $d$ is the dimensionality of the junction. While for 1D junctions Shapiro peaks should be observable at frequencies of tens of GHz and just fractions of a Kelvin away from a typical critical temperature $T_c \sim 10$ K, in order to produce reasonable sharp resonant features for a 3D junction one would need to use frequencies of at least hundreds of GHz -- several THz and be able to measure temperature with almost millivolt precision.

In order to take into account thermal (Gaussian) voltage fluctuations in the voltage-biased Josephson Junction we must add an additional noise term $v(t)$ to the applied voltage $V(t)$ with the following correlation function:
\begin{equation}
    \langle v(t) v(t')\rangle = 2RT\delta(t - t')\,,
\end{equation}
where $R$ is the tunneling resistance in normal state. The presence of the additional voltage noise term does not affect the main line of our calculations. It would appear in the expression for the dc current, Eq.~(\ref{J0tau}), as an additional factor $\exp\!\left[2\ii e\int_{t_1}^t\!\dd t'v(t')\right]$ in the integral. Averaging over this noise can be performed with the help of a cumulant expansion,

\begin{figure}[!tbp]
    \includegraphics[width=\columnwidth]{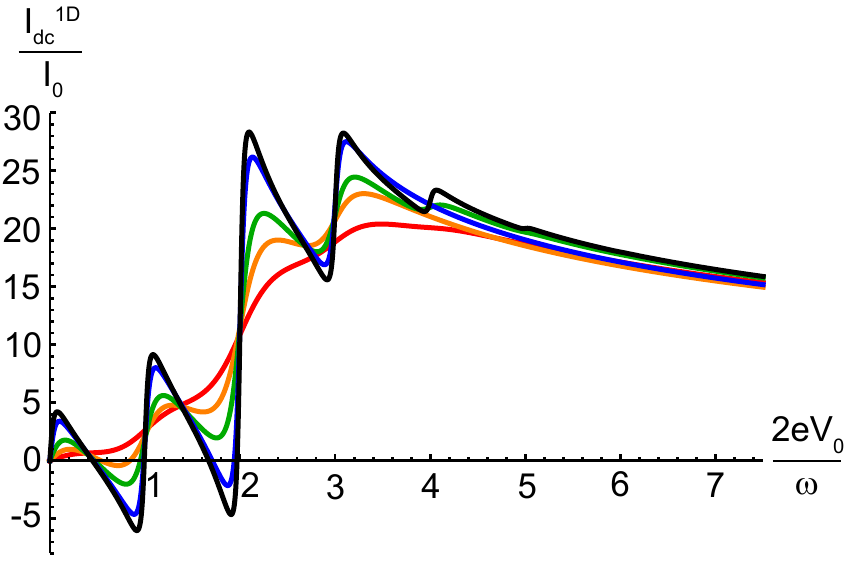}
    \caption{Normalized tunneling current of fluctuation Cooper pairs as a function of $V_0$ for various noise levels, $\Lambda =$ 0(black), 1(blue), 5(green), 10(orange), 20(red), where we used the same parameters as in Fig.~(\ref{J1D_J3D} a). White thermal voltage noise does not have any visible effect on the IV characteristic of 1D junctions at $\Lambda < 1$, while at the values of the noise level parameter $\Lambda > 1$ Shapiro resonances become smeared.}\label{noise}
\end{figure}

\begin{equation}
    \left\langle \ee^{2\ii e\int_{t_1}^t\!\dd t'v(t')}\right\rangle = \ee^{-2e^2\left\langle\left[\int_{t_1}^t\!\dd t'v(t')\right]^2\right\rangle} = \ee^{-4e^2RT(t - t_1)}\,,
\end{equation}
to yield the following final expression for the IV characteristics (1D case, assuming $\gamma_1 = \gamma_2 = 0$):
\begin{widetext}
\begin{align}\label{1D_noise}
    \langle I\rangle_\text{dc}^{1D} &= \frac{I_0}\epsilon \!\sum\limits_{k=-\infty}^{\infty}\!J_k^2\!\left(\frac{2eV_1}\omega\right)
    \int\limits_0^\infty\!\dd t'\, \sin\left[\left(\frac{2eV_0}\omega + k\right)\omega\tau_\text{GL}\,t'\right]
    \frac{\ee^{-(1 + 4e^2RT\tau_\text{GL})t'}}{\sqrt{t'}}\,\rm{erfc}\left(\sqrt{t'}\right)\,.
\end{align}
\end{widetext}
Numerical analysis based on Eq.~(\ref{1D_noise}) for a typical junction with normal state resistance $R \sim 10\Omega$ shows that the effects of voltage fluctuations only become noticeable ($\Lambda \equiv 4e^2RT\tau_\text{GL} \gtrsim~1$) at temperatures as close to critical as $\epsilon \lesssim 4 \times 10^{-3}$, see Fig.~(\ref{noise}). Even when $\Lambda \gtrsim 1$ the spikes in the IV characteristics become only mildly smeared, with the features vanishing at $\Lambda \gtrsim 10$. Therefore, the predicted effect of Shapiro spikes due to superconducting fluctuations should be experimentally observable in a wide range of temperatures.

To sum up, we have developed a simple phenomenological theory of resonant transport of fluctuation Cooper pairs through the tunneling barrier under the influence of external radiation. The above consideration generalizes the physics of the well known Shapiro effect for the temperature range above $T_c$, where this phenomenon can be used for determination of the Cooper pair lifetime.
Extension of Shapiro resonances onto the $T > T_c$ range, where they are caused by resonant tunneling of the \textit{fluctuation} Cooper pairs, opens yet one more area of applications of this phenomenon. Since tuning frequency ensures the best accuracy as compared to variation of other controlling parameters, the observation of the Shapiro spikes in the fluctuation regime can provide unprecedented precision measurements of the lifetime of fluctuation Cooper pairs. Hence, contrary to other fluctuation effects, Shapiro spikes grow even sharper upon approaching $T_c$, this precision is restricted only by the intrinsic noise of the system.

This work was supported by the U.S. Department of Energy, Office of Science, Materials Sciences and Engineering
Division. The work of ASM was also supported by the Russian Foundation for Basic Research and the grant of the
Russian Ministry of Science and Education $(02.\text{B}.49.21.0003)$.

\end{document}